\documentclass[pdflatex,sn-nature]{sn-jnl}
\usepackage{graphicx}%
\usepackage{multirow}%
\usepackage{amsmath,amssymb,amsfonts}%
\usepackage{amsthm}%
\usepackage{mathrsfs}%
\usepackage[title]{appendix}%
\usepackage{xcolor}%
\usepackage{textcomp}%
\usepackage{manyfoot}%
\usepackage{booktabs}%
\usepackage{algorithm}%
\usepackage{algorithmicx}%
\usepackage{algpseudocode}%
\usepackage{listings}%

\hypersetup{
  colorlinks = true,
  urlcolor   = blue,
  citecolor  = blue,
  linkcolor  = blue
}

\newcommand{\templatetype}[1]{}
\newcommand{\leadauthor}[1]{}

\newcommand{\showmatmethods}[1]{}
\newcommand{\showacknow}[1]{}
\newcommand{\dates}[1]{\date{#1}}

\makeatletter
\def\storedsignificance{}
\def\storedauthorcontributions{}
\def\storedauthordeclaration{}
\def\storedcorrespondingauthor{}
\def\storedmethods{}
\def\storeddataavailability{}
\def\storedacknowledgements{}

\long\def\authorcontributions#1{\gdef\storedauthorcontributions{#1}}
\long\def\authordeclaration#1{\gdef\storedauthordeclaration{#1}}
\long\def\correspondingauthor#1{\gdef\storedcorrespondingauthor{#1}}
\long\def\matmethods#1{\gdef\storedmethods{#1}}

\newcommand{\printstoredfrontmatter}{%
  \ifx\storedsignificance\@empty\else
    \section*{Significance statement}
    \storedsignificance
  \fi
}

\newcommand{\printstoredbackmatter}{%
  \ifx\storedauthorcontributions\@empty\else
    \section*{Author contributions}
    \storedauthorcontributions
  \fi
  \ifx\storedauthordeclaration\@empty\else
    \section*{Competing interests}
    \storedauthordeclaration
  \fi
  \ifx\storedcorrespondingauthor\@empty\else
    \section*{Correspondence}
    \storedcorrespondingauthor
  \fi
\ifx\storedmethods\@empty\else
  \section*{Methods}
  \storedmethods
\fi
  \ifx\storeddataavailability\@empty\else
    \section*{Data availability}
    \storeddataavailability
  \fi
  \ifx\storedacknowledgements\@empty\else
    \section*{Acknowledgements}
    \storedacknowledgements
  \fi
}
\makeatother

\graphicspath{{FIGURES/}{Figures/}}

\DeclareMathOperator{\pe}{\mathit{Pe}}

\DeclareMathOperator{\nus}{\mathit{Nu}}

\DeclareMathOperator{\ra}{\mathit{Ra}}

\title[Universal transport laws in buoyancy-driven porous mixing]{Universal transport laws in buoyancy-driven porous mixing}

\author*[1]{\fnm{Marco} \sur{De Paoli}}\email{marco.de.paoli@tuwien.ac.at}
\author*[2]{\fnm{Xiaojue} \sur{Zhu}}\email{zhux@mps.mpg.de}

\affil[1]{\orgdiv{Institute of Fluid Mechanics and Heat Transfer}, \orgname{TU Wien}, \orgaddress{\postcode{1060}, \city{Vienna}, \country{Austria}}}
\affil[2]{\orgname{Max Planck Institute for Solar System Research}, \orgaddress{\postcode{37077}, \city{G\"ottingen}, \country{Germany}}}

\leadauthor{De Paoli}

\authorcontributions{Author contributions: M.D. and X.Z. designed research, performed research, analyzed data, acquired funding and wrote the paper.}

\authordeclaration{The authors declare no conflict of interest.}

\correspondingauthor{\textsuperscript{1}To whom correspondence should be addressed. E-mail: marco.de.paoli@tuwien.ac.at; \textsuperscript{2}E-mail: zhux@mps.mpg.de}


\abstract{
Buoyancy-driven convection in porous media governs heat and mass transport in a wide range of natural and engineered systems, from groundwater aquifers and geothermal reservoirs to carbon storage in geological formations and flows through planetary interiors. Yet the transient regime, in which fingering flows emerge and transport is strongly enhanced, is still described largely through empirical scaling laws, limiting predictive capability across conditions. Here we show that transient porous buoyancy-driven mixing obeys exact time-dependent balances that couple transport, flow intensity, and scalar dissipation. These balances remain accurate when restricted to the actively mixing layer, revealing that the essential dynamics are localized within a finite region. Leveraging these results, we derive a minimal one-parameter closure for the mean scalar field. The theory we propose predicts self-similar mean profiles, universal second-order statistics, and a linear transport law without case-by-case tuning. Direct numerical simulations up to spatial resolution of $2048\times2048\times16384$ points validate these predictions. Our results place transient porous mixing on a predictive footing, showing how macroscopic transport laws emerge from exact balances and self-similar dynamics, and provide a general framework for buoyancy-driven transport in porous media.
}

\matmethods{
\subsection*{Numerical Simulations}

The domain spans $L^*_x,L^*_y$ in the horizontal directions ($x^*, y^*$), while $L^*_z$ is the domain size in vertical direction $z^*$, along which gravity $\mathbf{g}$ acts.
Here $^*$ is used to indicate dimensional variables, and the absence of superscript refers to dimensionless variables.
The flow variables are scaled according to convective scales as in \cite{pir21}. Specifically, the velocity field $\mathbf{u^*}=(u^*,v^*,w^*)$ is scaled with $\mathcal{U}^* =  g \Delta\rho^* K / \mu$. 
Lengths are scaled by the domain height $L^*_z$, and time $t^*$ is scaled with $\phi L^*_z/\mathcal{U}^*$.
The dimensionless temperature, $T$, is obtained as $T = (T^* - T^*_\text{mean}) / \Delta$, where $T^*$ is the dimensional temperature field and $T^*_\text{mean}=(T^*_\text{max}+T^*_\text{min})/2$ is the mean temperature.
The pressure scale is set by $g L^*_z(t^*) [\rho^*(T^*_\text{min}) - \rho^*(T^*_\text{max})]$, using the initial fluid density difference across the top-to-bottom layers, $\rho^*(T^*_\text{min}) - \rho^*(T^*_\text{max})$.
We assume that the Oberbeck-Boussinesq approximation applies.
The domain is periodic in horizontal direction and closed in vertical direction, i.e., there is no flux of mass and heat across the horizontal boundaries ($w^*=0$, $\partial_{z^*}T^*=0$ at $z^*=\pm L_z^*/2$).
The flow is initially at rest, $\mathbf{u}^*(t^*=0)=0$, and characterized by a step-like temperature field (see Fig.~\ref{fig:intro}a), such that $T^*(z^*>0)=T^*_\text{min}$ and $T^*(z^*<0)=T^*_\text{max}$ (despite our focus being on thermally driven flows in a porous medium in thermal equilibrium with the fluid, similar insights apply to solute-driven flows as long as $\ra$ is the same).

In this incompressible system, heat transport is governed by the dimensionless advection-diffusion equation \citep{pir21}:
\begin{eqnarray}
    \frac{\partial T}{\partial t} + \mathbf{u} \cdot \nabla T = \frac{1}{\ra} \nabla^2 T,
    \label{eq:01}
\end{eqnarray}
where $\mathbf{u}$ and $T$ denote the velocity and temperature fields, respectively.
Momentum transport and flow incompressibility are represented by the Darcy law and continuity equation:
\begin{equation}
    \mathbf{u} = -(\nabla p - T \mathbf{k}),
    \label{eq:03}
\end{equation}
\begin{equation}
    \nabla \cdot \mathbf{u} = 0,
    \label{eq:04}
\end{equation}
where $p$ is the reduced pressure field and $\mathbf{k}$ is the unit vector in the $z$ direction.
As time progresses, sufficiently high values of $\ra$ give rise to distinct columnar finger-like structures that develop across the domain in both 2D and 3D, with vertical mixing marking the flow's dynamic progression (Fig.~\ref{fig:intro}b-ii).
The mixing length is defined as $h$ and evolves as a function of time $t$.
The evolution of $h(t)$ thus introduces the time dependency to $\ra$, which dynamically reflects the growing scale of the mixing region in the porous medium (see Fig.~\ref{fig:intro}b-ii).
A quantitative definition for $h$ is presented in Supplementary information.

The governing equations, namely Eqs.~\eqref{eq:01}, \eqref{eq:03} and \eqref{eq:04}, are solved numerically with the open-source second-order finite difference solver AFiD-Darcy \citep{depaoli2025code}, previously employed to simulate convective porous media flows \citep{depaoli2025dispersion,depaoli2025afid,depaoli2025morphology}. 
The code consists of an efficient parallel solver for simulating convective, incompressible, and wall-bounded flows in porous media. The algorithm employs a pressure-correction scheme in combination with an efficient Fast Fourier Transform-based solver. Additional details on the algorithm and the initial condition are described in the Supplementary information.
A list of the simulations performed and relative details is reported in Tab.~\ref{tab:numdet_text}.
}

\dates{This manuscript was compiled on \today}

\begin{document}
\maketitle

\printstoredfrontmatter

\begin{figure}
\centering
\includegraphics[width=0.99\columnwidth]{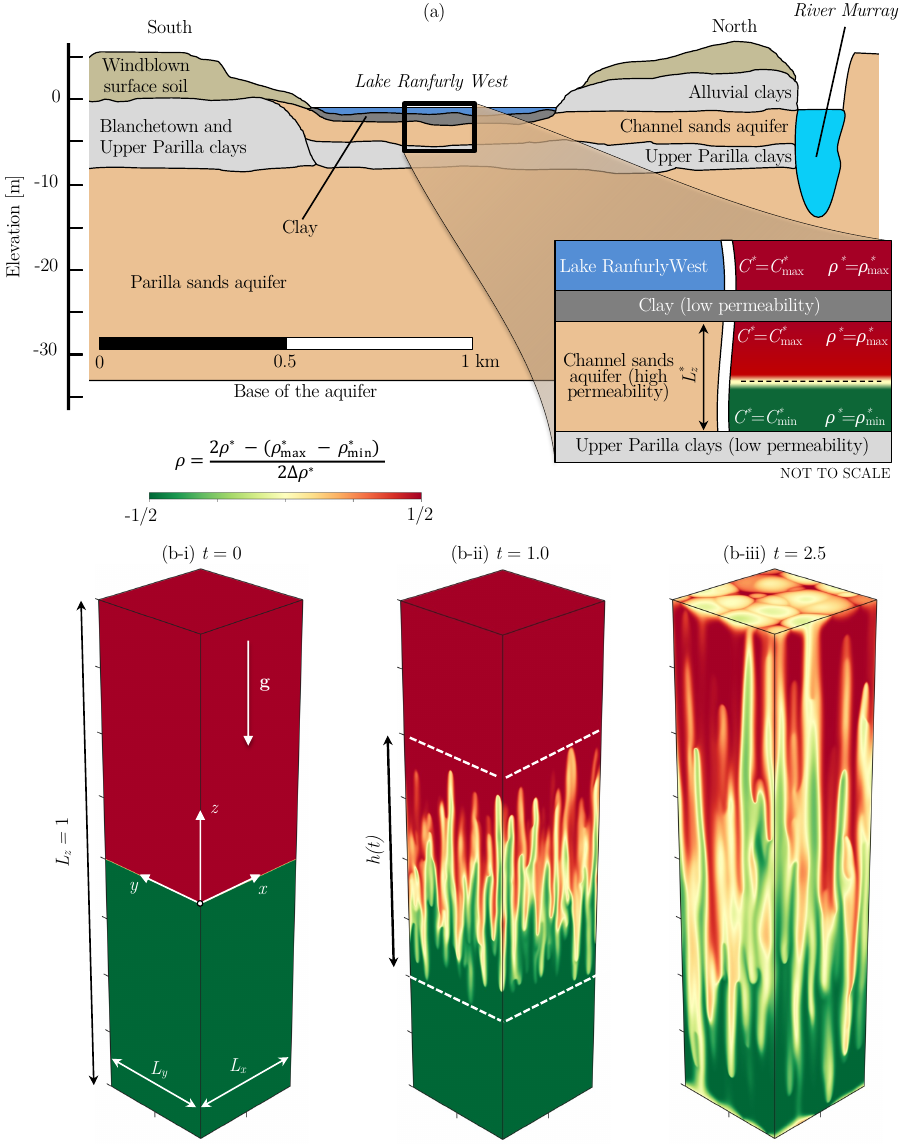}
\caption{
(a) Conceptualized hydrogeology of the River Murray basin area (New South Wales, Australia), adapted from Narayan et al.~\cite{narayan1995simulation}.
Inset: modelling of the saline seepage through the bottom of Lake Ranfurly West. The high-permeability sands aquifer is confined by two low-permeability layers. Lake Ranfurly West supplies high-salt-concentration (high-density) water from the top, while low-salinity (low-density) groundwater is present in the aquifer.
(b)~Evolution of the density distribution in time ($t$) for simulation S1 ($\ra=3.2\times10^4$, quantities shown in dimensionless units).
(b-i)~Initial flow configuration with indication of gravity ($\mathbf{g}$), of the reference frame ($x,y,z$) and of the domain extension ($L_x,L_y,L_z$).
(b-ii)~A developed field during the convective phase, with indication of the extension of the mixing region, $h$.
(b-iii)~Field after the fingers reached the boundaries.}
\label{fig:intro}
\end{figure}

Transport in porous media shapes the evolution of many natural and engineered systems, from groundwater aquifers and geothermal reservoirs to geological carbon storage and hydrothermal activity in planetary interiors \citep{cheng1979heat,jolie2021geological,heinze2023velocity,ulloa2025convection,goodman2004hydrothermal,lowell2005hydrothermal,le2020internally,schrag2007preparing,orr2009onshore,szulczewski2012lifetime}. In these settings, buoyancy can destabilize an initially stratified fluid and drive vigorous convective mixing, greatly accelerating the transport of heat and dissolved species. In contrast, diffusive spreading reduces local density differences, and thus the strength of this driving mechanism. 
This competition between convection and diffusion also underlies solute mixing in geological formations \citep{,narayan1995simulation,depaoli2025simulation} and salt seepage from saline lakes \cite{narayan1995simulation,depaoli2025dispersion}. An example of the latter is reported in Fig.~\ref{fig:intro}(a), where the hydrogeology of the Murray River basin is schematically illustrated. The high-salinity water channeled to the Lake Ranfurly West seeps through the low-permeability lake bottom lake into the underneath Channel Sands Aquifer, connected to the Murray River. In these permeable sands, a high-salinity fluid layer builds up, resulting in an unstable configuration with a heavy fluid that sits above the lighter low-salinity groundwater. Eventually, when the density contrast between the two fluid layers is large enough, convection sets in. Predicting the dynamics of this convective mixing is key to evaluate the influx on salt into the adjacent Murray River. Yet, despite the key importance of this process, its transient dynamics remain poorly understood at a predictive level.

A central difficulty is that transient buoyancy-driven mixing is usually described through empirical scaling laws. These are useful within the range of conditions for which they were obtained, but they offer limited confidence when extrapolated across the broad parameter regimes relevant to geophysics and engineering. What is still missing is a framework in which transport laws emerge from the dynamics themselves, rather than being inferred case by case from simulation or experiment.

In porous materials, this problem arises naturally in the Rayleigh-Taylor-Darcy configuration, where a denser fluid overlies a lighter one in a permeable matrix \citep{rayleigh1883investigation,taylor1950instability}. In subsurface settings, the motion is controlled by Darcy's law, and the resulting instability generates descending and rising fingers, an expanding mixing layer, and transport rates far larger than diffusion alone. Previous studies have established a robust phenomenology, including approximately linear growth of the mixing region \citep{DeWit2004,gopalakrishnan2017relative,Boffetta2020}, finite-size corrections \citep{depaoli2019universal,depaoli2019prf,depaoli2022experimental}, and strongly enhanced transport at large driving \citep{hew12,hew14,wen2013computational,wen2015structure,pir21,Boffetta2020}. But these observations do not yet amount to a predictive theory: the dominant transport laws and leading statistics of the mixing process are still not derived from exact time-dependent constraints.

This gap is especially striking when compared with statistically steady porous convection, for which exact relations have recently clarified how global transport is linked to flow intensity and dissipation \citep{zhu2024transport}. In that setting, the heat-transfer rate is known to increase linearly with the strength of buoyancy forcing in the strongly convective regime, and this result can be derived from exact balances rather than inferred empirically, in the spirit of Grossmann-Lohse theory \citep{grossmann2000scaling,gro01,lohse2024ultimate}. No comparable theoretical structure has yet been established for transient porous Rayleigh-Taylor mixing. Existing descriptions typically treat the mixing-layer thickness as an external ingredient or rely on empirical closures \citep{Boffetta2020,borgnino2021dimensional,boffetta2022dimensional}, leaving open the deeper question of whether transient porous mixing also obeys universal balances that can predict both structure and transport.

Here we show that it does. We derive exact time-dependent budget identities for transient porous Rayleigh-Taylor mixing and demonstrate that, during the convection-dominated regime, these balances remain accurate when restricted to the actively mixing layer. Leveraging  these results, we develop a minimal one-parameter closure for the horizontally averaged scalar field, building on the nonlinear eddy-diffusivity framework of Boffetta et al.~\cite{boffetta2010prl} and adapting it to Darcy dynamics. The resulting theory predicts self-similar mean profiles, universal second-order statistics, and the associated transport law, all of which are validated by three-dimensional direct numerical simulations over a broad range of Rayleigh-Darcy numbers.

\section*{Results}\label{sec:rel}
We consider a homogeneous and isotropic porous medium having height $L_z^*$ and permeability $K$, which is initially saturated with a light fluid at bottom (temperature $T^*=T^*_\text{max}$, density $\rho^* = \rho^*_\text{min}$) and a cold, heavy fluid at top ($T^*=T^*_\text{min}$, $\rho^*=\rho^*_\text{max}$).
The initial temperature-induced density difference defines an unstable configuration, due to the presence of gravity $g$ acting in vertical direction.
A sketch of the flow configuration is provided in Fig.~\ref{fig:intro}(b).
To quantify the efficiency of convective transport, we measure the Nusselt number $\nus$, defined as the total vertical heat flux normalized by its purely diffusive value.
Note that the problem formulation is independent of the transported scalar considered, i.e., whether heat or mass fluxes are considered.
We derive the results in the frame of thermal convection in analogy with previous works \citep{boffetta2010prl}.
However, our findings are valid also for mass transport settings as we show in the Discussion section. 
The strength of the buoyancy forcing is characterized by the Rayleigh-Darcy number $\ra = g \Delta\rho^* K L_z^*/(\kappa \mu)$
which compares buoyancy-driven advection with diffusive damping. Here, $\Delta\rho^*=\rho^*_\text{max}-\rho^*_\text{min}$ is the initial density, $\kappa$ is the thermal diffusivity, and $\mu$ is the dynamic viscosity.
We simulate the evolution for $3.2\times10^4\le \ra \le 2.56\times10^5$, with the highest-$Ra$ case resolved on a $2048\times2048\times16384$ grid, marking the largest simulation in this configuration ever. 
We used the superscript $^*$ to indicate dimensional fields and dimensionless fields (no superscript) are obtained as discussed in the Methods section.

\subsection*{Exact global relations}\label{sec:rela}
Integrating the advection-diffusion equation~(\ref{eq:01}) over the domain yields
$d \langle T \rangle/{d t } = 0$,
where $\langle\cdot\rangle$ denotes volume average, showing conservation of the mean temperature.

Taking the dot product of Darcy's law (\ref{eq:03}) with ${\bf u}$ and averaging gives
$\left\langle |{\bf{u}}|^2 \right\rangle
= -\left\langle {\bf{u}} \cdot \nabla p \right\rangle
+ \left\langle w T \right\rangle$.
Using incompressibility and the divergence theorem, and under periodic or no-flux boundary conditions, the pressure term vanishes
\begin{equation}
\left\langle |{\bf{u}}|^2 \right\rangle= \left\langle w T \right\rangle.
\label{eq:peb2}
\end{equation}
Multiplying by $\ra^2$ yields
$\pe^2 = \left\langle w T \right\rangle\ra^2$,
where $\pe=\mathcal{V}\ra$ is the Pèclet number, with $\mathcal{V}=\sqrt{\langle|{\bf u}|^2\rangle}$.
With the Nusselt number definition
$\nus = 1 + \ra \left\langle w T \right\rangle$,
we obtain the exact relation
\begin{equation}
\pe^2 = (\nus - 1)\,\ra,
\label{eq:bud1}
\end{equation}
which links convective transport to flow intensity in a parameter-free way \citep{hassanzadeh2014wall,zhu2024transport}, showing that transport is slaved to flow intensity.

\begin{figure}
    \centering
    \includegraphics[width=0.75\linewidth]{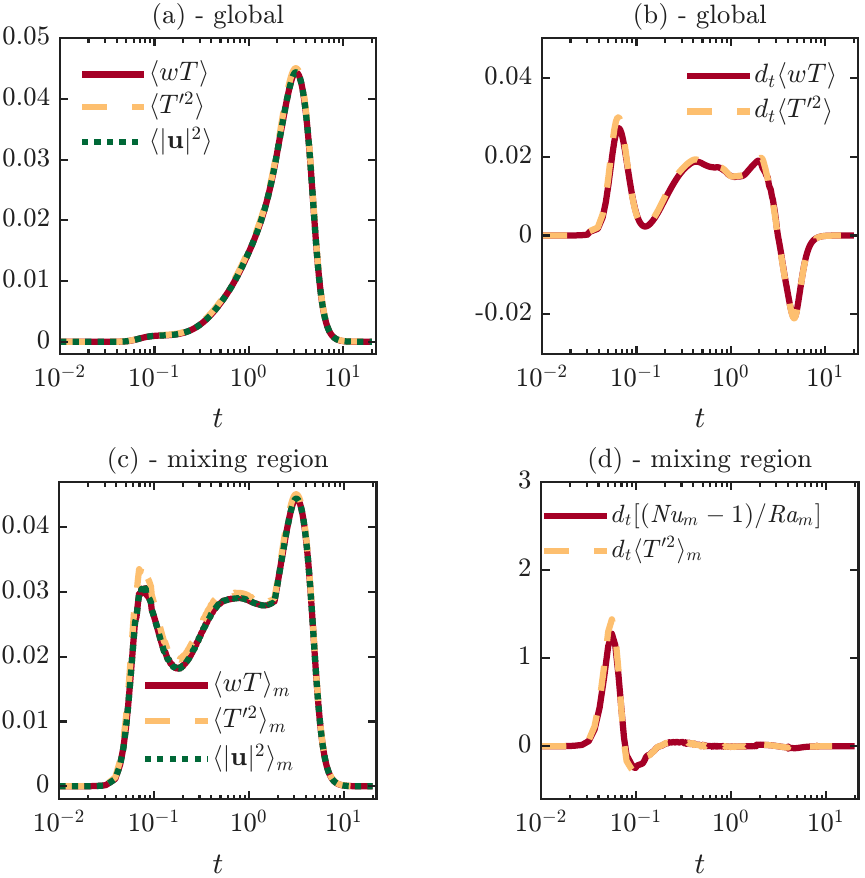}
    \caption{
    Verification of the thermal fluctuations relations against numerical simulations ($\ra=3.2\times10^4$, corresponding to simulation S1, see Tab.~\ref{tab:numdet_text} for further details).
    The validity of Eqs.~\eqref{eq:peb2} and \eqref{eq:ref32} and the validity of Eqs.~\eqref{eq:ref32cd} is verified in (a) and (b), respectively.
    The validity of Eqs.~\eqref{eq:peb2b} and \eqref{eq:ref32b} and the validity of Eqs.~\eqref{eq:budgetb2} is verified in (c) and (d), respectively.
    }
    \label{fig:mixtherm}
\end{figure}

To characterize thermal mixing, we consider the evolution of the volume-averaged temperature variance. Multiplying Eq.~\eqref{eq:01} by $T$ and integrating over the domain gives (derivation in Supplementary information)
\begin{equation}
\frac{1}{2}\frac{d\langle T^2 \rangle}{dt}  = -\chi,
\label{eq:bud2}
\end{equation}
where $\chi=\langle |{\nabla T}|^2 \rangle/\ra$ is the mean scalar dissipation.

Finally, decomposing $T=\overline{T}+T'$ and using the mid-plane symmetry of the Rayleigh-Taylor configuration yields the exact fluctuation relation (derivation in Supplementary information)
\begin{equation}
\langle w T\rangle =\langle T'^2\rangle.
\label{eq:ref32}
\end{equation}
The predictions provided by Eqs.~\eqref{eq:peb2} and \eqref{eq:ref32} are verified to be robust throughout the entire mixing process, as shown in Fig.~\ref{fig:mixtherm}(a) where their validity is verified against numerical simulations.
The relations are key as they allow one to connect flow intensity, convective transport and temperature fluctuations.

Taking the derivative in time of Eq.~\eqref{eq:ref32}, we obtain
\begin{equation}
\frac{d\langle w T\rangle}{dt} =\frac{d\langle T'^2\rangle}{dt},
\label{eq:ref32cd}
\end{equation}
and using Eq.~\eqref{eq:bud2}, one gets:
\begin{equation}
\frac{1}{\ra}\frac{d \nus }{dt}=-2\chi-\frac{d\langle \overline{T}^2 \rangle}{dt},
\label{eq:budget}
\end{equation}
which provides an exact global relation quantifying the evolution of dissipative and transport mechanisms in RTD flows, and is verified against numerical simulations in Fig.~\ref{fig:mixtherm}(b).

\subsection*{Approximate relations within the mixing region}\label{sec:relb}
The mixing layer is the region populated by the fingers as indicated in Fig.~\ref{fig:intro}(b-ii).
Its vertical extension $h(t)$ is computed here using the fitting procedure proposed by Boffetta et al.~\cite{boffetta2010prl} and based on the horizontally averaged temperature profile (see Supplementary information for details and comparisons with alternative definitions). We then treat the mixing layer as an approximately ``closed'' region during the convection-dominated stage, i.e. assuming $w\simeq 0$ and $\partial_z T\simeq 0$ at its horizontal boundaries, located at  $z = \pm  h/2$.
These approximations hold after the initial diffusive stage, when no finger has formed yet, and until the fingers touch the horizontal boundaries, when confinement effects enter the picture. 
Quantities defined within the mixing region are labeled with the subscript $m$.

Repeating the previous procedure over the mixing-layer volume yields identities analogous to the domain-wide relations.
Defining $\ra_m=\ra h$ and $\pe_m=\mathcal{V}_m\ra_m$ with $\mathcal{V}_m=\sqrt{\langle |{\bf u}|^2\rangle_m}$, we obtain
\begin{equation}
\langle\left | \mathbf{u} \right |^2 \rangle_m=\langle T'^2\rangle_m,
\label{eq:peb2b}
\end{equation} and $\pe_m^2 = \left\langle w T \right\rangle_m\ra_m^2$.
With $
\nus_m = 1 + \ra_m \left\langle w T \right\rangle_m$ it follows that
\begin{equation}
\pe_m^2 = (\nus_m - 1)\ra_m.
\label{eq:bud1b}
\end{equation}
Similarly, the mixing-layer variance budget reads
\begin{equation}
\frac{1}{2}\frac{d\langle T^2 \rangle_m}{dt}  = -\chi_m,
\label{eq:bud2z}
\end{equation}
where $\chi_m=\langle |{\nabla T}|^2 \rangle_m/\ra$.
Repeating the reasoning relative to the global symmetry argument (see Supplementary information) yields:
\begin{equation}
\langle w T\rangle_m =\langle T'^2\rangle_m.
\label{eq:ref32b}
\end{equation}
The relations (\ref{eq:peb2b})-(\ref{eq:ref32b}) are not exact, but DNS show they become precise once convection is established.
For instance, the excellent accuracy provided by Eqs.~\eqref{eq:peb2b} and \eqref{eq:ref32b} is verified in Fig.~\ref{fig:mixtherm}(c).
In addition, they support the interpretation that the essential dynamics are localized within the fingered region (additional discussions on the quantitative diagnostics are available in Supplementary information).

Taking the time derivative of Eq.~\eqref{eq:ref32b} we obtain
\begin{equation}
\frac{d}{dt}\left(\frac{\nus_m-1}{\ra_m}\right)=\frac{d\langle T'^2 \rangle_m}{dt},
\label{eq:budgetb2}
\end{equation}
which, combined with Eq.~\eqref{eq:bud2z}, leads to the diagnostic mixing-layer budget
\begin{equation}
\frac{d}{dt}\left(\frac{\nus_m-1}{\ra_m}\right)
=-2\chi_m-\frac{d\langle \overline{T}^2 \rangle_m}{dt}.
\label{eq:budgetb}
\end{equation}
Equation~(\ref{eq:budgetb}) connects the instantaneous evolution of the mixing-layer transport to dissipation and to the evolution of the horizontally averaged temperature fluctuations.
Equation~(\ref{eq:budgetb2}) is used as mixing-layer diagnostics (its validity is verified in Fig.~\ref{fig:mixtherm}d) and as motivation for the reduced mixing-layer closure developed in the next section.

In summary, the global identities provide exact constraints linking transport, flow intensity, and dissipation in a parameter-free way, while their mixing-layer counterparts (though approximate) show that the essential dynamics are largely confined to the fingered zone. This motivates the reduced description adopted next, where the mixing layer is treated as a single evolving region and modeled through a minimal eddy-diffusivity closure.

\begin{figure}
    \centering
    \includegraphics[width=0.85\linewidth]{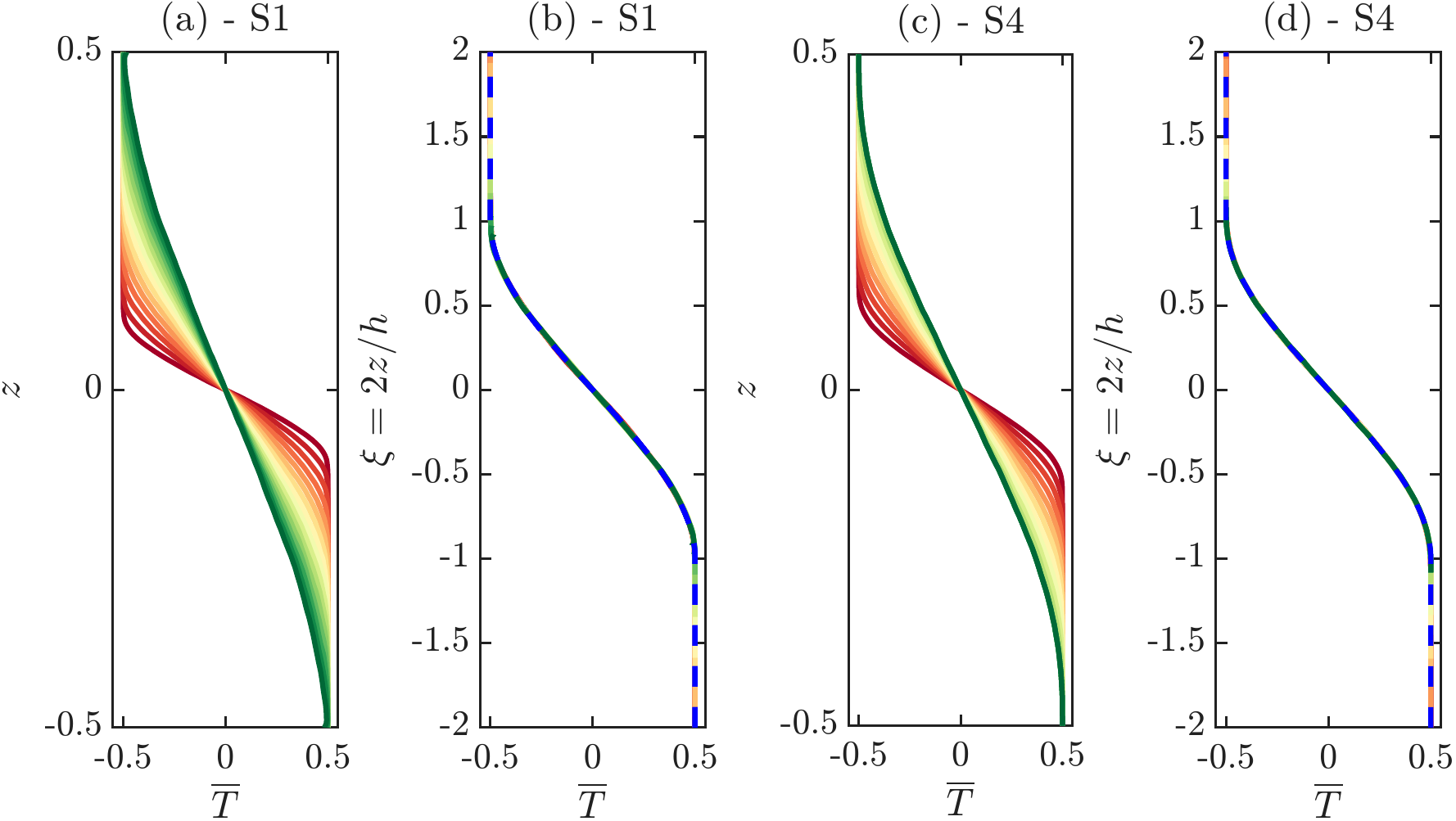}
    \includegraphics[width=0.75\linewidth]{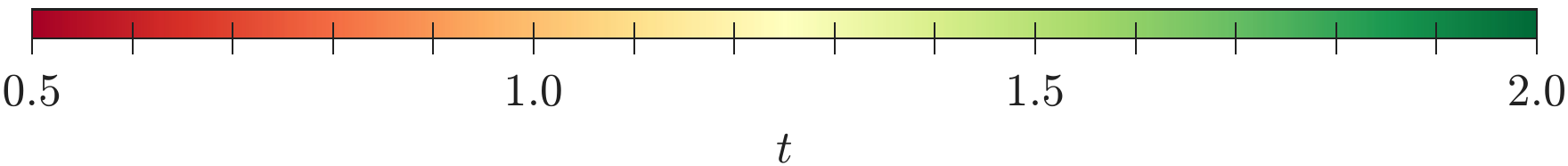}
    \caption{
    Profiles obtained at $\ra=3.2\times 10^4$ (simulation S1, panels a and b) and $\ra=2.56\times 10^5$ (simulation S4, panels c and d) for $0.5\le t \le 2.0$.
    Original profiles (panels~a,c) and profiles rescaled using the self-similar coordinate $\xi$ (panels~b,d) are shown.
    The blue dashed line represents the self-similar solution derived in Eq.~\eqref{eq:cases}.
    }
    \label{fig:profiles}
\end{figure}

\begin{figure}
    \centering
    \includegraphics[width=0.65\linewidth]{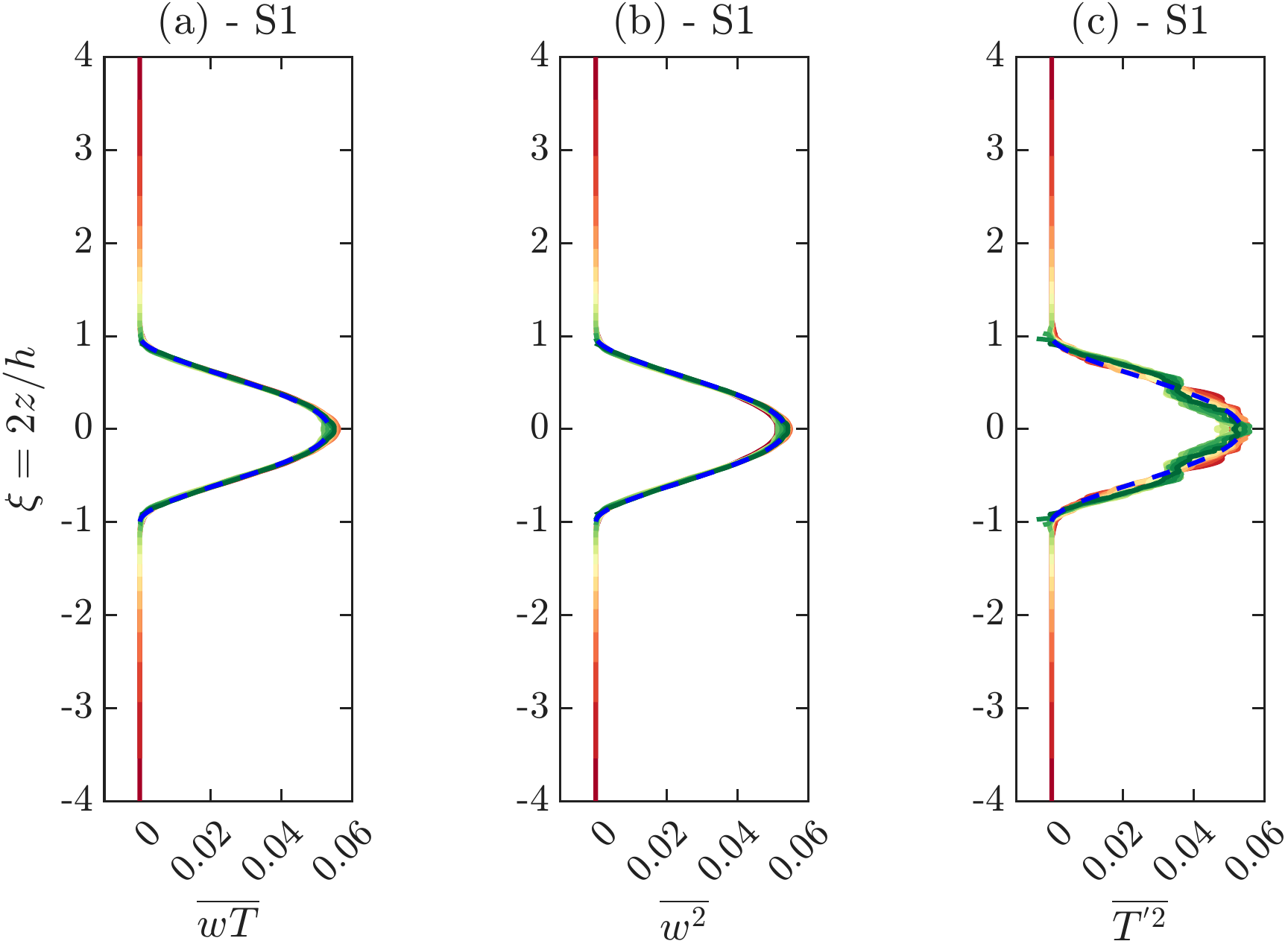}\\
    \vspace{0.2cm}\includegraphics[width=0.65\linewidth]{Figures/colormap.pdf}\vspace{0.2cm}
    \includegraphics[width=0.65\linewidth]{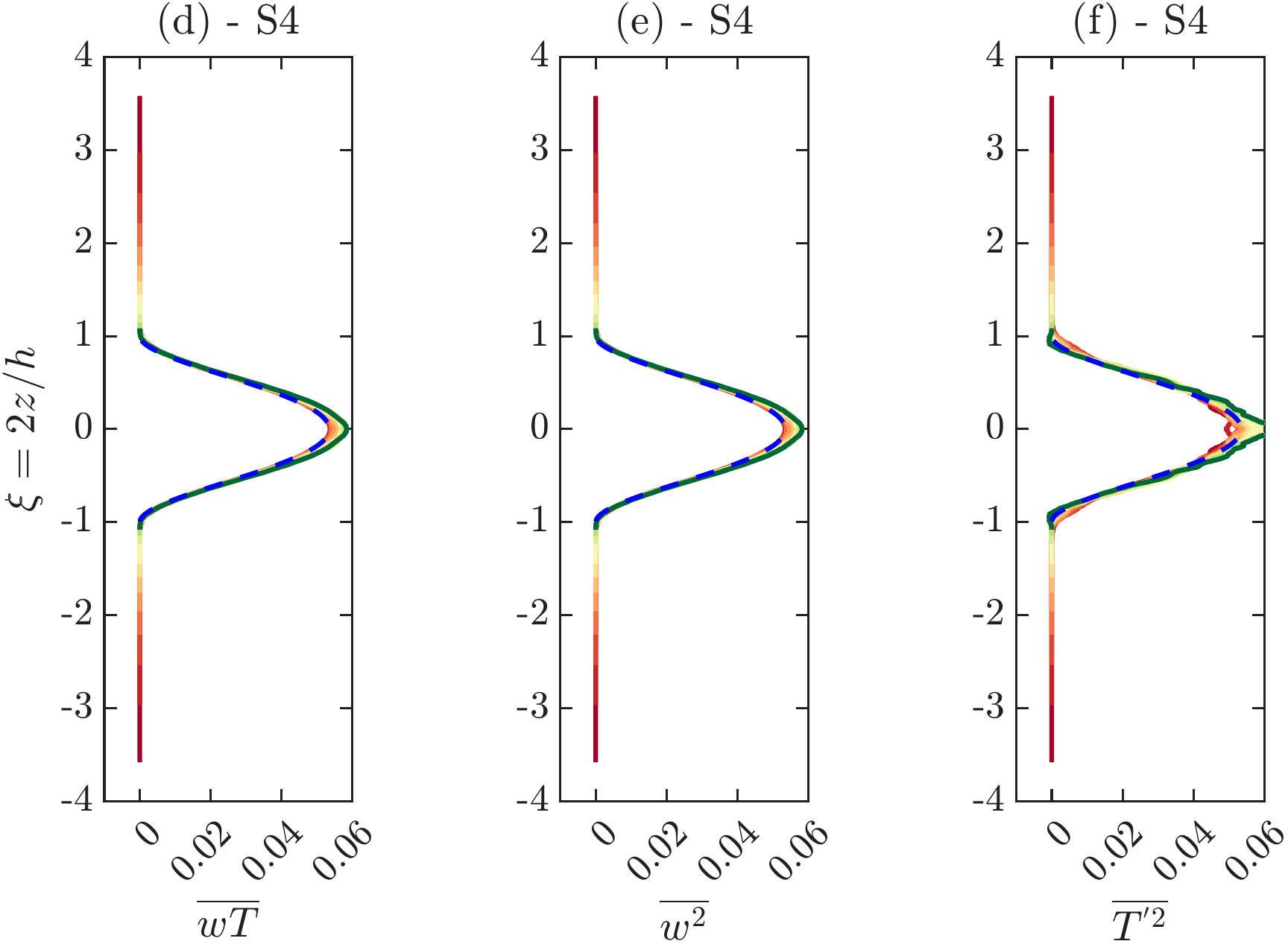}
    \caption{Verification of Eq.~\eqref{eq:w2bis}, with the blue dashed line given by $9a/16(1-\xi^2)^2$. 
    Simulations S1 ($\ra=3.2\times10^4$) and S4 ($\ra=2.56\times10^5$) are considered in the top and bottom panels, respectively. 
    Additional cases are reported in Supplementary information.}
    \label{fig:corre45}
\end{figure}

\subsection*{Eddy-diffusivity closure model for mean profile}

The global and mixing-layer relations show that the fingered region behaves, to leading order, as a quasi-closed active layer whose evolution controls transport.
This suggests that a reduced model could be built for the horizontally-averaged scalar field, using the mixing-layer thickness as the dominant evolving length scale.
We therefore introduce a minimal eddy-diffusivity closure for the convective flux and derive its predictions for profiles and transport.

Following Boffetta et al.~\cite{boffetta2010prl}, we write the convective flux as
\begin{equation}
\overline{wT}=-K(z,t)\,\partial_z\overline{T},
\label{eq:42}
\end{equation}
and assume a gradient-dependent eddy diffusivity,
\begin{equation}
K(z,t)=a\,z_1^2(t)\,|\partial_z\overline{T}|,
\label{eq:43}
\end{equation}
where $z_1(t)=h(t)/2$ is the mixing-layer half-thickness and $a$ is a dimensionless coefficient setting the mixing amplitude.
The physical motivation for this model is that mixing is not homogeneous within the mixing layer, since temperature and velocity fluctuations decrease moving away from the origin (see also Fig.~\ref{fig:corre45} and Supplementary information).
The model~(\ref{eq:43}) reflects the mixing length theory proposed by Prandtl \cite{boffetta2010prl,siggia1994high,prandtl1925bericht}, in which the eddy diffusivity can be expressed as $K(z,t)\sim\ell^2\,\partial_z \Phi$ where $\ell$ represents a characteristic length of the mixing region and $\Phi$ is the typical fluctuation of the field considered (temperature in this case).

Horizontally averaging Eq.~\eqref{eq:01} yields
\begin{equation}
\frac{\partial \overline{T}}{\partial t}+\frac{\partial}{\partial z}\overline{wT}
=\frac{1}{\ra}\frac{\partial^2\overline{T}}{\partial z^2}.
\label{eq:avgheat}
\end{equation}
In the convection-dominated stage we neglect molecular diffusion within the mixing layer, and substituting Eqs.~\eqref{eq:42} and \eqref{eq:43} into \eqref{eq:avgheat} we obtain the nonlinear diffusion equation
\begin{equation}
\frac{\partial\overline{T}}{\partial t}
=\frac{\partial}{\partial z}\!\left[a\,z_1^2(t)\,(\partial_z\overline{T})\,|\partial_z\overline{T}|\right].
\label{eq:nonlin_diff_meanT}
\end{equation}
Inside the mixing region, we seek a solution $\Theta(\xi)$ function of the self-similar coordinate $\xi$ such that:
\begin{equation}
\overline{T}(z,t)=\Theta(\xi),\qquad \xi=\frac{z}{z_1(t)},\qquad |\xi|\le 1,
\label{eq:simform}
\end{equation}
with $\overline{T}=\pm 1/2$ outside the mixing region ($|\xi|>1$) and assuming anti-symmetry with respect to $z=0$, i.e., $\Theta(-\xi)=-\Theta(\xi)$.
At the mixing-layer edges ($\xi=\pm1$) we impose zero flux and matching to the outer states,
\begin{equation}
\partial_z\overline{T}(\pm z_1,t)=0
\quad\text{and}\quad
\overline{T}(\pm z_1,t)=\mp\frac12.
\label{eq:BC_mixing_layer}
\end{equation}
Using the similarity solution function described in Eq.~\eqref{eq:simform} into Eq.~\eqref{eq:nonlin_diff_meanT}, the problem reduces to an ODE of the form
\begin{equation}
\dot z_1\,\xi=2a\,\Theta''(\xi).
\label{eq:sep_variables_appendix}
\end{equation}
The left-hand side depends on $t$ (through $\dot{z}_1$) and on $\xi$, while the right-hand side depends only on $\xi$.
It follows that for a solution to exist over a finite time interval, $\dot{z}_1$ must be constant, and thus $\dot z_1=U$ is constant over the self-similar stage, so that $z_1(t)=U(t-t_0)$, being $t_0$ an integration constant.
Integrating twice and enforcing the boundary conditions (\ref{eq:BC_mixing_layer}) yields $U=3a$ and therefore the linear growth law
\begin{equation}
z_1(t)=3a\,(t-t_0),
\label{eq:z1final}
\end{equation}
together with the cubic similarity profile
\begin{equation}
\overline{T}(z,t)=
\begin{cases}
\dfrac12, & \xi\le -1,\\[3pt]
-\dfrac34\,\xi+\dfrac14\,\xi^3, & |\xi|<1,\\[3pt]
-\dfrac12, & \xi\ge 1.
\end{cases}
\label{eq:cases}
\end{equation}
This solution represents an excellent approximation to the profiles obtained numerically.
We report in Fig.~\ref{fig:profiles} two examples relative to simulations S1 (Fig.~\ref{fig:profiles}a and Fig.~\ref{fig:profiles}b) and S4 (Fig.~\ref{fig:profiles}c and Fig.~\ref{fig:profiles}d).
When rescaled using the self-similar coordinate $\xi$, the profiles collapse on the solution proposed in Eq.~\eqref{eq:cases}, highlighting the self-similar character of the flow during the convective phase.

The cubic shape derived in Eq.~\eqref{eq:cases} coincides with the self-similar mean profile obtained by Boffetta et al.~\cite{boffetta2010prl} for Rayleigh-Taylor turbulence in clear fluids, but the growth law differs: in the Navier--Stokes case $z_1\propto t^2$, whereas here Darcy dynamics enforce $z_1\propto t$.
Consequently, the transport scaling differs between the two systems, and will be analyzed in detail in the following.

\subsection*{Energy transport in the mixing region}
Within the mixing layer, the flux profile predicted by Eq.~\eqref{eq:42} using the self-similar solution (\ref{eq:cases}) gives 
\begin{equation}
\overline {wT}= \frac{9a}{16}(1-\xi^2)^2,
\quad |\xi|\le1,
\label{eq:411}
\end{equation}
and the corresponding mean mixing-layer flux is
\begin{equation}
\langle wT\rangle_m = \frac{3a}{10}.
\label{eq:mixfluxmean}
\end{equation}

Guided by the exact global balances derived ($\langle |{\bf u}|^2\rangle=\langle wT\rangle=\langle T'^2\rangle$), by the Darcy law~\eqref{eq:03} itself and by the fingering-dominated structure of the mixing layer, we expect temperature and vertical velocity to be strongly correlated, with $|{\bf u}|^2\simeq w^2$ and similar vertical structures for $w^2$, $T'^2$, and $wT$. We therefore adopt the minimal closure
\begin{equation}
\overline{w^2} \simeq \overline{T'^2} \simeq \overline{wT}=\frac{9a}{16}(1-\xi^2)^2,\quad |\xi|\le1.
\label{eq:w2bis}
\end{equation}
Equation (\ref{eq:w2bis}) is well supported by the DNS results over the convection-dominated stage.
The results reported in Fig.~\ref{fig:corre45} are relative to simulations S1 and S4 (additional data available in Supplementary information), and indicate an excellent predictive capability of the model proposed, that with just one coefficient, $a$, allows to accurately describe the self-similar evolution of these statistics during the entire convective regime.
The best-fitting parameter $a$ is reported for each simulation in Tab.~\ref{tab:numdet_text} and appears to be universal, i.e., independent of $\ra$.

\begin{figure}
    \centering
    \includegraphics[width=0.75\linewidth]{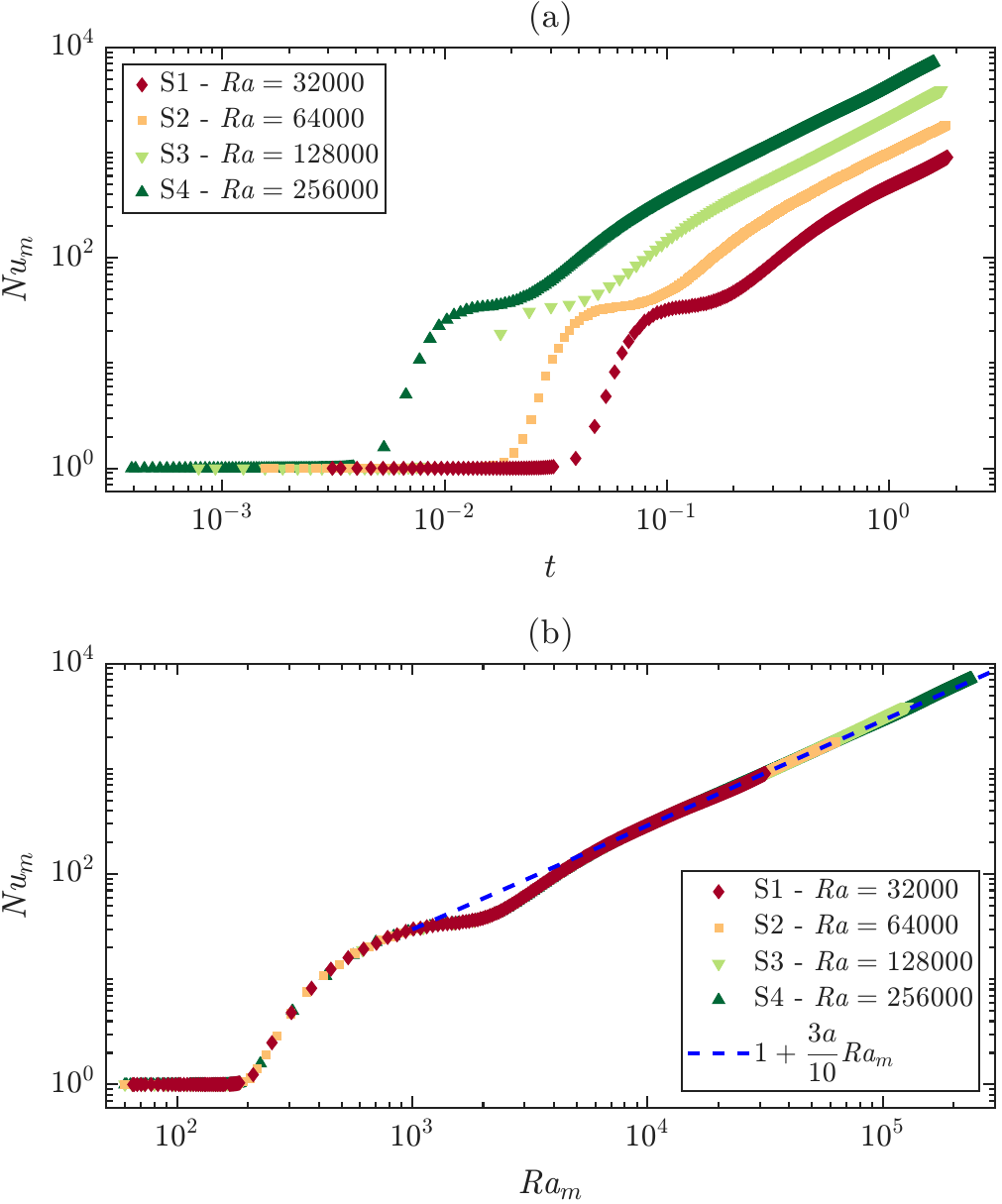}
    \caption{
    Evolution of the mixing-layer Nusselt number $\nus_m$ in time~(a) and as a function of the instantaneous Rayleigh number $\ra_m$~(b) for the simulations performed (symbols).
    Equation (\ref{eq:numml2}) (dashed line in panel b) provides an excellent description of the collapsed transport data, using a single coefficient, $a$ (details in Supplementary information).
    }
    \label{fig:nussra}
\end{figure}

Averaging Eq.~\eqref{eq:w2bis} over the mixing layer yields
\begin{equation}
\langle w^2\rangle_m = \langle T'^2\rangle_m = \langle wT\rangle_m = \frac{3a}{10}.
\label{eq:mix34}
\end{equation}
Since $h=2z_1$, Eq.~\eqref{eq:z1final} implies a linear growth of the local Rayleigh number,
\begin{equation}
\frac{\ra_m}{\ra}=h=6a(t-t_0).
\label{eq:linfh}
\end{equation}
Using the definition $\nus_m = 1 + \ra_m \left\langle w T \right\rangle_m$ and Eq.~\eqref{eq:mixfluxmean}, we obtain the predicted mixing-layer transport law
\begin{equation}
\nus_m = 1+\frac{3a}{10}\ra_m.
\label{eq:numml2}
\end{equation}
Although $\nus_m(t)$ differs across Rayleigh numbers (Fig.~\ref{fig:nussra}a), a self-similar behavior is once again observed for $\nus_m(\ra_m)$, with all the data relative to different simulations collapsing onto the predicted linear trend during the convection-dominated stage (Fig.~\ref{fig:nussra}b).
Additional details to verify the model's ability to capture the long-term behavior are provided in Supplementary information.

In summary, the proposed eddy-diffusivity model provides a compact, self-consistent description of the RTD mixing-layer structure and transport in terms of a single evolving length scale, $z_1(t)$, and one dimensionless coefficient $a$.
Once calibrated, the model captures the collapsed mean profile, the leading fluctuation organization, and the linear mixing-layer transport law (\ref{eq:numml2}).
Simulations confirm these findings and provide conclusive evidence that the self-similar solution (\ref{eq:cases}) represents an accurate description of the flow evolution. 

\section*{Discussion}\label{sec:conc}
We show that porous media convection is not as unconstrained as it appears. Transient Rayleigh-Taylor-Darcy mixing obeys exact, time-dependent budget identities that tightly bound transport, flow intensity, and scalar fluctuations. When restricted to the evolving mixing layer, these constraints remain accurate throughout the convection-dominated regime, revealing that the essential dynamics are localized within a finite active region. In this framework, we introduced a minimal eddy-diffusivity closure for the heat flux as in Eq.~\eqref{eq:42} that reduces the problem to a single evolving length scale and one universal dimensionless coefficient. This theory unifies what are often treated as separate questions -- structure, mixing intensity and flux -- into one minimal and yet physically-sound description that requires no case-by-case tuning. In contrast to empirical approaches, transport laws emerge directly from exact balances and self-similar dynamics. High-resolution three-dimensional DNS confirm the universality of this organization across a wide range of Rayleigh-Darcy numbers, and up to unprecedented values.

Present findings can be used, for instance, to predict the long-term spreading of contaminants in groundwater flows. 
With respect to the dispersion of salt in the Murray River basin, schematically illustrated in Fig.~\ref{fig:intro} and described in Narayan et al.~\cite{narayan1995simulation}, we have that the typical density contrast between the concentrated and fresh layers, which drives the flow, is $\Delta\rho^*=52.5$~kg~m$^{-3}$, and we consider the viscosity ($\mu=10^{-3}$~Pa~s) and the solute diffusivity ($D=1.5\times10^{-9}$~m$^2$~s$^{-1}$) constant and independent of the salt concentration.
The region between the low permeability layers, modeled in the inset of Fig.~\ref{fig:intro}(a), is confined from above by the bottom of the lake and from below by the Upper Parilla clays, and it extends vertically for $L_z^*\approx4$~m.
This layer is characterized by a permeability $K=2.95\times10^{-11}$~m$^2$ and porosity $\phi=0.3$.
It follows that the convective velocity is $\mathcal{U}^*=g\Delta\rho^*K/\mu=1.52\times10^{-5}$~m~s$^{-1}=1.31$~m~d$^{-1}$, leading to a Rayleigh-Darcy number $\ra=\mathcal{U}^*L_z^*(\phi D)^{-1}\approx1.35\times10^5$ and a characteristic time $\mathcal{T}^*= \phi L^*_z(\mathcal{U}^*)^{-1}=0.91$~d.
Using Eq.~\eqref{eq:linfh} we obtain that the extension of the mixing region evolves as:
\begin{equation}
    h^* = h\;L^*_z = 6\;\mathcal{U}^*a(t^*-t_0\mathcal{T}^*)\approx 6\;\mathcal{U}^*a\; t^* \approx 0.756 \text{ m d}^{-1} \times t^*.
    \label{eq:applic1}
\end{equation}
This simple calculation indicates that it takes about $t^*_b\approx L^*_z/0.756\text{ m d$^{-1}$}\approx5.3$~days for the fingers to extend up to the horizontal low-permeability boundaries.
We can estimate the amount (expressed in kg~m$^{-2}$s$^{-1}$) of salt mixed due to convection as:
\begin{equation}
    J^*_\text{conv} = \frac{\nus_m D \Delta C^*}{h^*} \approx \frac{3 a\;\mathcal{U}^* \Delta C^*}{10\phi},
    \label{eq:applc3}
\end{equation}
where $\Delta C^* = 84$~kg~m$^{-3}$.
It follows that $J^*_\text{conv}$ is constant and equal to 10.5~kg~m$^{-2}$~d$^{-1}$.
In the absence of convection, and then with an initially steady and flat interface between the two layers, the diffusive mixing rate evolves in time as
\begin{equation}
    J^*_\text{diff} = \Delta C^* \sqrt{\frac{D}{\pi t^*}}.
    \label{eq:applc4}
\end{equation}
For comparison, we computed its value at $t^*_b$, which gives $J^*_\text{diff}(t^*_b) =0.23$~kg~m$^{-2}$~d$^{-1}$, which is about 50 times smaller then in the corresponding convective scenario.

\clearpage

\backmatter

\printstoredbackmatter

\bmhead{Acknowledgments}{We gratefully acknowledge the financial support from the Max Planck Society, the German Research Foundation through grants 521319293, 540422505, 550262949, the Alexander von Humboldt Foundation, and the Daimler Benz Foundation. The project is also funded by the European Union (ERC, MORPHOS, 101163625). Views and opinions expressed are however those of the author(s) only and do not necessarily reflect those of the European Union or the European Research Council. Neither the European Union nor the granting authority can be held responsible for them. The computational results presented have been achieved using the Vienna Scientific Cluster (VSC).}

\begin{table}[t!]
\centering
\caption{\label{tab:numdet_text}
Numerical details of the simulations performed.
Simulation number (S\#) and Rayleigh-Darcy number ($\ra$) are reported.
A uniform grid is employed, with resolution $N_x \times N_y \times N_z$.
The largest simulation (S4), in which $\ra$ is doubled with respect to S3, is performed on a domain with $L_x=L_y=1/8$ to keep the computational cost affordable.
Finally, the best fitting values of the coefficient $a$ are reported, with a mean value $a=0.0960$.
}
\begin{tabular}{cccccc}
Sim. & $\ra$ & $L_x=L_y$ & $L_z$ & $(N_x \times N_y) \times N_z$ & $a$\\
\midrule
S1 & $3.2 \times 10^4$ & 1/4 & 1 & $512^2 \times 2048$    & 0.0957\\
S2 & $6.4 \times 10^4$ & 1/4 & 1 & $1024^2 \times 4096$   & 0.0934\\
S3 & $1.28 \times 10^5$ & 1/4 & 1 & $2048^2 \times 8192$  & 0.0974\\
S4 & $2.56 \times 10^5$ & 1/8 & 1 & $2048^2 \times 16384$ & 0.0975\\
\bottomrule
\end{tabular}
\end{table}

\bibliography{sn-bibliography}

\end{document}